\documentclass[]{emulateapj}

\slugcomment{Accepted for publication in ApJ, Feb}

\shorttitle{NIR RADIAL VELOCITIES WITH AN AMMONIA CELL}   %limited to 44 characters
\shortauthors{Bean et al.}

\begin{document}

\title{The CRIRES Search for Planets Around the Lowest-Mass Stars. \\
       I.~High-Precision Near-Infrared Radial Velocities with an Ammonia Gas Cell$^{1}$}

\author{
Jacob L.~Bean\altaffilmark{2,7},
Andreas Seifahrt\altaffilmark{2,3},
Henrik Hartman\altaffilmark{4},
Hampus Nilsson\altaffilmark{4},
G\"unter Wiedemann\altaffilmark{5},
Ansgar Reiners\altaffilmark{2,8},
Stefan Dreizler\altaffilmark{2},
\& Todd J.~Henry\altaffilmark{6}
}

\email{bean@astro.physik.uni-goettingen.de}

\altaffiltext{1}{Based on observations made with ESO Telescopes at the Paranal Observatories under program ID 182.C-0748}

\altaffiltext{2}{Institut f\"ur Astrophysik, Georg-August-Universit\"at,
  Friedrich-Hund-Platz 1, 37077 G\"ottingen, Germany}

\altaffiltext{3}{Department of Physics, University of California, One Shields Avenue, Davis, CA 95616, USA}

\altaffiltext{4}{Lund Observatory, Lund University, Box 43, 22100 Lund, Sweden}

\altaffiltext{5}{Hamburger Sternwarte, Gojenbergsweg 112, 21029 Hamburg, Germany}

\altaffiltext{6}{Department of Physics and Astronomy, Georgia State University, Atlanta, GA, 30302, USA}

\altaffiltext{7}{Marie Curie International Incoming Fellow}

\altaffiltext{8}{Emmy Noether Fellow}

\begin{abstract}
Radial velocities measured from near-infrared spectra are a potentially powerful tool to search for planets around cool stars and sub-stellar objects. However, no technique currently exists that yields near-infrared radial velocity precision comparable to that routinely obtained in the visible. We are carrying out a near-infrared radial velocity planet search program targeting a sample of the lowest-mass M dwarfs using the CRIRES instrument on the VLT. In this first paper in a planned series about the project, we describe a method for measuring high-precision relative radial velocities of these stars from $K$-band spectra. The method makes use of a glass cell filled with ammonia gas to calibrate the spectrograph response similar to the ``iodine cell'' technique that has been used very successfully in the visible. Stellar spectra are obtained through the ammonia cell and modeled as the product of a Doppler-shifted template spectrum of the object and a spectrum of the cell, convolved with a variable instrumental profile model. A complicating factor is that a significant number of telluric absorption lines are present in the spectral regions containing useful stellar and ammonia lines. The telluric lines are modeled simultaneously as well using spectrum synthesis with a time-resolved model of the atmosphere over the observatory. The free parameters in the complete model are the wavelength scale of the spectrum, the instrumental profile, adjustments to the water and methane abundances in the atmospheric model, telluric spectrum Doppler shift, and stellar Doppler shift. Tests of the method based on the analysis of hundreds of spectra obtained for late M dwarfs over six months demonstrate that precisions of $\sim$\,5\,m\,s$^{-1}$ are obtainable over long timescales, and precisions of better than 3\,m\,s$^{-1}$ can be obtained over timescales up to a week. The obtained precision is comparable to the predicted photon-limited errors, but primarily limited over long timescales by the imperfect modeling of the telluric lines.
\end{abstract}


\section{INTRODUCTION}
\subsection{Motivation for NIR radial velocities}
The vast majority of known exoplanets have been detected with high-precision (i.e. $\sigma$ $<$ 10\,m\,s$^{-1}$) time-series stellar radial velocity measurements. Invariably, these measurements have been made using spectra obtained in the visible wavelength region. This is natural for two reasons. First, arguably the most interesting targets for planet searches are solar-type stars (broadly defined here as dwarfs with F, G, and K spectral types), and these stars are brightest at wavelengths shorter than 1\,$\mu$m. The spectra of solar-type stars at these wavelengths are also rich in deep and sharp spectral lines suitable for Doppler shift measurements. The second reason is that visible wavelength spectrograph technology is much more advanced relative to that required for spectrographs operating in other wavelength regions. When fed by a moderately sized telescope, visible wavelength spectrographs using a cross-dispersed echelle design and CCD detectors can yield high-resolution and high signal-to-noise (S/N) spectra with a large wavelength coverage for a large number of solar-type stars. The more advanced technology in the visible also includes well-established methods for high-precision calibration. The combination of currently obtainable data quality and calibration together make it possible to reach radial velocity precisions of $\sim$\,1\,m\,s$^{-1}$ for many solar-type stars, which is sufficient to detect planets down to a few Earth masses in short-period orbits around them \citep[e.g.][]{howard09,bouchy09}.

Despite the obvious utility of radial velocity measurements in the visible, measurements at other wavelengths are also desirable. The near-infrared (NIR, i.e. 1 -- 5\,$\mu$m) in particular is an interesting wavelength region for radial velocity exoplanet studies. In contrast to solar-type stars, low-mass objects that are cooler than $\sim$\,4000\,K are brightest at wavelengths of 1\,$\mu$m and longer. Furthermore, stars cooler than $\sim$\,3200\,K are so faint at visible wavelengths that high-precision radial velocity measurements are impossible for all but a few very nearby examples even with the best instruments on 8-10\,m class telescopes. For example, the sample of 40 stars in a long-term radial velocity planet search program specifically targeting low-mass stars that utilized UVES \citep{dekker00} at the VLT only included two objects with masses below 0.2\,M$_{\sun}$, and only three with masses below 0.35\,M$_{\sun}$ \citep{zechmeister09}. 

Very low-mass stars are an interesting sample of potential planet hosts notwithstanding their neglected status. They are useful for probing the correlation between gas giant planet frequency and stellar mass \citep{endl06, johnson07}, are the most numerous stars in the Galaxy \citep{reid02, henry06, covey08}, and exhibit a larger dynamical response to orbiting planets and have closer-in habitable zones than higher-mass stars \citep{kasting93}. The latter point means that much less precision is needed to detect potentially habitable planets around low-mass stars with the radial velocity method than is needed for solar-type stars. The ubiquity of mid to late M dwarfs and the closeness of their habitable zones suggests that the closest stars hosting transiting habitable planets are likely of this stellar type \citep{deming09}. Furthermore, \citet{deming09} have estimated that basic atmospheric characterization of a few such planets could be obtained with the \textit{James Webb Space Telescope} using the techniques of transit and occultation spectroscopy. Therefore, very low-mass stars are an important sample for the possible future study of the atmosphere of a habitable planet. However, we first must be able to use the radial velocity method to identify or confirm (in the case the planet is first identified by a transit detection), and measure the masses of low-mass planets around very low-mass stars before these investigations can be carried out.

As faint as the lowest-mass stars are at visible wavelengths, their extreme redness means they are reasonably bright at NIR wavelengths. For example, an M6 star (M$_{\star}$ $\approx$ 0.1\,M$_{\sun}$) at 10\,pc will have a $V$-band magnitude of $\sim$\,15.5, which would make it a very challenging target for high-precision radial velocity measurements in this wavelength region with current telescopes and instruments. However, the same star will have a $K$-band magnitude of $\sim$\,8.5, which is bright enough that spectra with S/N suitable for high-precision radial velocity measurements could be easily obtained using existing high-resolution NIR spectrographs on 8-10\,m class telescopes - especially those equipped with adaptive optics (AO) systems. Therefore, if the issues and advantages of low-mass stars described above are to be studied and exploited using the radial velocity method, then the NIR spectral region offers the only possible option without building larger telescopes.

Radial velocities measured from NIR spectra also likely offer the advantage that activity induced ``jitter'' is reduced relative to measurements in the visible. Magnetic activity plays a role in limiting the effectiveness of radial velocity planet searches because it results in inhomogeneities on the stellar surfaces (spots and plage). Visible spectra are thus modulated by stellar rotation and variations in the activity. These intrinsic variations mimic real radial velocity changes and are very confusing for planet searches \citep[e.g.][]{queloz01,paulson04,wright05}. It is thought that radial velocities measured from NIR spectra will exhibit reduced jitter relative to the visible because of the smaller contrast between the cool inhomogeneities caused by activity and the average stellar surface at the longer wavelengths. Significant evidence exists to support the existence of this effect, and this evidence is discussed in the following sub-section.

The issue of activity induced radial velocity jitter is particularly relevant for low-mass stars because it is well established that a much higher fraction of late-type stars are active than earlier-type stars. In the most comprehensive survey so far, \citet{west04} found that the fraction of active stars increased from $<$ 10\% at spectral type M3, to 50\% at M5, and ultimately 75\% at the end of the main sequence. As high as these activity fractions are, they are likely only lower limits for nearby stars because the \citet{west04} survey included stars orbiting perpendicular to the Galactic plane, and such stars are potentially very old. Therefore, the fraction of active stars in a radial velocity planet search of M dwarfs will likely be even higher than the \citet{west04} statistics suggest because they will be drawn from an intrinsically younger sample. This has serious implications for planet searches because it means that the majority of the lowest-mass stars will exhibit substantial activity induced jitter in radial velocities measured from visible wavelength spectra even if their faintness can be overcome. 

\subsection{Previous work in the area}
The advantages offered by NIR radial velocities have been identified before and there has been some previous work to realize them. \citet{martin06} used NIRSPEC \citep{mclean98} on Keck to make NIR radial velocity measurements of the brown dwarf LP\,944-20. Their measurements over six nights had an rms dispersion of 360\,m\,s$^{-1}$. This result was inconsistent with previously measured visible wavelength radial velocities, which exhibited a coherent periodic signal with an amplitude of 3.5\,km\,s$^{-1}$. This result is generally interpreted as confirmation that radial velocity jitter is indeed reduced at NIR wavelengths for active cool dwarfs. However, it should be noted that the NIR and visible data sets were not contemporaneous, and the possibility that the star was experiencing a period of reduced activity during the NIR measurements can not be ruled out. 

\citet{zapatero09} have also reported radial velocity measurements obtained from NIRSPEC data, in this case for the very low-mass star VB\,10. Their data were obtained over seven years and have an estimated precision of $\sim$\,300\,m\,s$^{-1}$. These measurements seem to support the astrometric detection of a giant planet around VB\,10 claimed by \citet{pravdo09}. However, our measurements of VB\,10, which are presented in paper II and exhibit a dispersion of only 10\,m\,s$^{-1}$, are inconsistent with the \citet{zapatero09} measurements and also appear to completely rule out the existence of the proposed giant planet around this star \citep{bean10}. In addition, \citet{anglada10} did not detect the expected reflex motion of VB\,10 in their visible wavelength radial velocities, which have a typical precision $\sim$\,200\,m\,s$^{-1}$.

\citet{blake07} reported NIR radial velocity measurements with PHOENIX \citep{hinkle03} on Gemini South. They targeted L dwarfs and reached a precision of 300\,m\,s$^{-1}$ over five nights. \citet{prato08} observed a sample of young stars and some late-type radial velocity standard stars with CSHELL \citep{greene93} on the IRTF for seven nights. They reached precisions of $\sim$\,100\,m\,s$^{-1}$ for the standard stars, and 100 -- 300\,m\,s$^{-1}$ for the young stars. The measurements for the young stars again indicated no coherence with significant periodicities seen in visible radial velocities, but also were not contemporaneous. Similarly, \citet{huelamo08} monitored the young star TW Hya over six nights using CRIRES \citep{kaeufl04} on the VLT. Their measurements yielded an rms dispersion of 35\,m\,s$^{-1}$, which was inconsistent with a coherent signal having an amplitude of $\sim$\,300\,m\,s$^{-1}$ seen in visible radial velocities obtained previously \citep{setiawan08} and contemporaneously. \citet{seifahrt08} also used CRIRES to obtain continuous observations of an M giant over five hours and reached a precision of $\sim$\,20\,m\,s$^{-1}$.

In addition to these stellar observations, there has been some work on measuring NIR radial velocities for the Sun. \citet{deming87} and \citet{deming94} presented data from a long-term monitoring program using a Fourier Transform Spectrometer (FTS). Their $K$-band measurements had a precision of better than 5\,m\,s$^{-1}$. More recently, \citet{ramsey08} used a laboratory grating spectrograph to measure the signature of the Earth's rotational velocity in $Y$-band spectra of integrated Sunlight. They maintained a precision of $\sim$\,10\,m\,s$^{-1}$ over a few hours in two separate experiments. 

\subsection{Challenges of NIR radial velocities}
Despite the attention paid to obtaining NIR radial velocities of cool stars, no previous work has achieved a long-term precision on a star other than the Sun within an order of magnitude of the precision that is routinely obtained in the visible. One reason for this is simply that enough signal was not obtained to probe the true limit in some cases. However, based on photon statistics, the observations of \citet{prato08} and \citet{huelamo08} should have at least achieved precisions around a factor of two better than they did. Another issue is that no studies with an external check on the obtained precision have been reported continuing for longer than seven days. Therefore, the obtainable long-term precision in NIR radial velocities has not been systematically developed or evaluated. 

The main issue that has limited the pursuit of high-precision NIR radial velocities is the lack of a suitable calibration method. For example, useful lines from available emission lamps \citep[e.g. ThAR,][]{lovis07} are infrequent in the NIR relative to the visible. This is a problem for the existing high-resolution NIR spectrographs because they generally have relatively short wavelength coverage for single exposures. Laser frequency combs have the potential to provide calibration lines in nearly any spectral region conceivable for stellar radial velocity work \citep{murphy07,li08,steinmetz08}. However, the technique is still not developed enough for long-term use at an astronomical observatory. Furthermore, none of the existing NIR spectrographs can be fed simultaneously with light from a calibration source like a lamp or laser comb system, and are not stabilized to the level necessary to do without simultaneous calibration. In addition, there is also the issue that current NIR spectrographs experience significant variations in illumination on short timescales. This means that even if these instruments could be simultaneously fed with light from a lamp, the calibration would not track all the necessary effects for high-precision radial velocity measurements. Therefore, emission lamps and laser combs are not useful for high-precision calibration in currently available NIR spectrographs, but could be for future instruments designed for larger wavelength coverage and excellent stability \citep{reiners10}.

All of the previous studies discussed above that targeted stars other than the Sun utilized the telluric spectrum imprinted on the data for \textit{in situ} calibration, although the lines and methods utilized varied somewhat. In their study, \citet{seifahrt08} demonstrated that telluric N$_{2}$O lines near 4.1\,$\mu$m were stable to a level of 10\,m\,s$^{-1}$ over five hours. On the other hand, \citet{deming87} showed that the telluric methane lines in the $K$-band varied with a semi-amplitude of 20\,m\,s$^{-1}$ depending on the hour angle due to the prevailing winds over the observatory. In addition to the possible variability of the telluric lines, another issue with using telluric lines as a radial velocity fiducial is that measurements obtained over a full observing season could be affected by systematics because the stellar lines will move relative to the telluric lines due to the changing barycentric velocity of the observatory along the line of sight. This will lead to more or less blending of features, which could influence the measurements. 

The well known ``iodine cell'' method \citep{butler96} has been used to measure radial velocities to precisions of a few m\,s$^{-1}$ with numerous visible spectrographs that are not highly stabilized. The general gas cell technique involves placing a cell in front of the spectrograph during observations so that a standard spectrum is imprinted on each obtained stellar spectrum. The cell's spectrum recorded during each exposure can then be used to precisely determine the spectrograph response at the exact moment of the observations and, thus, calibrate the simultaneously obtained stellar spectra for radial velocity measurements.

The success of the iodine cell in the visible implies that a gas cell could be a useful way to calibrate currently available NIR spectrographs. The main constraint for the gas cell method is that the gas in question should exhibit numerous sharp and deep lines in a wavelength interval where the stars targeted for radial velocity measurements have lines as well \citep{campbell79}. Ideally, this region would also be free of atmospheric lines. However, this is a particularly challenging requirement in the NIR due to the prevalence of strong atmospheric lines even in the traditional windows between water absorption bands. 

\citet{deming87} and \citet{deming94} used a cell filled with N$_{2}$O to calibrate their FTS measurements of integrated sunlight from 2.0 -- 2.5\,$\mu$m. In addition, \citet{mahadevan09} considered a number of options for NIR gas cells and concluded that the gases H$^{13}$C$^{14}$N, $^{12}$C$_{2}$H$_{2}$, $^{12}$CO, and $^{13}$CO together could provide useful calibration in the $H$-band. However, there have been no reported uses of a gas cell for high-precision NIR radial velocity measurements for stars other than the Sun, and no such measurements obtained using a grating spectrograph. 

We have recently developed a new gas cell for simultaneous calibration of spectra obtained in the $K$-band, and are currently using it to conduct a radial velocity search for planets around very low-mass stars with the CRIRES instrument at the VLT. In this paper, which is the first in a planned series about our planet search, we describe a method for measuring high-precision relative radial velocities of cool stars using observations made with the cell, and we present extensive data sets demonstrating the performance of the method over long timescales. The paper is laid out as follows. In \S2 we describe the new cell and its implementation in CRIRES. We describe the typical observing procedure and data reduction used in \S3. We detail the radial velocity measurement algorithm in \S4. In \S5, we present tests of the obtained radial velocity precision. We conclude in \S6 with a discussion and outlook for future planet searches utilizing NIR radial velocities. 

\section{AMMONIA CELL}
The cell that we have developed for measuring high-precision NIR radial velocities contains ammonia ($^{14}$NH$_{3}$). Ammonia is a well established wavelength standard for the NIR \citep[e.g.][and references therein]{urban89}, but to our knowledge has not been used for calibration in astronomical observations. At room temperature ammonia is in its gaseous state and exhibits a rich molecular spectrum in the NIR even with the relatively low column densities possible in a cell to be used at an astronomical observatory. 

Although ammonia exhibits lines in different regions of the NIR, existing high-resolution NIR spectrographs yield relatively small wavelength coverage in a single exposure. Therefore, a careful consideration of the region for observations was necessary in parallel with the choice of a gas for the calibration cell. We chose to make observations in a window in the $K$-band because the very low-mass stars we are interested in exhibit numerous sharp and deep lines in this wavelength region suitable for radial velocity measurements (mainly from the first overtone transitions of CO), and ammonia exhibits a number of lines useful for calibration. The availability of a window that can be spanned by existing NIR spectrographs and that contains a significant number of both calibration lines and stellar lines is a particular advantage of ammonia over other possible gases, including those discussed in \S1.3.

Unfortunately, the window we are observing in also contains a significant number of absorption lines arising from telluric methane and water. This was known before we began our work. High-precision radial velocity measurements are usually made by avoiding regions containing telluric lines due to the expectation that these lines will exhibit variability on the order of a few to tens of m\,s\,$^{-1}$. However, as discussed in \S1, the lack of an obvious method for calibrating existing NIR spectrographs means that a more flexible approach is currently called for. We decided that using an ammonia cell in the $K$-band and accepting the contamination from the telluric lines was a good option considering all the competing issues.

The main body of the ammonia cell we are using is a glass tube that has a length of 17\,cm, and diameter of 5\,cm. The windows were chemically fused on to the ends of the tube. They are made of quartz glass with a low OH content and have excellent visible and NIR transmission (the visible transmission is important for the use of an AO system). Each window has a few degree wedge angle on one face to eliminate possible fringing when illuminated. The windows' wedges are oriented 180$\degr$ relative to each other to minimize the tilt induced on light passing through the cell. The cell was evacuated and filled with 50\,mb ammonia at 15$\degr$\,C via a small tube in the side of the body. This tube was heated and pinched off while maintaining a constant pressure. In principle, the bonds holding the windows on to the body of the cell should remain sealed for 10+ years. 

We have implemented the ammonia cell in the high-resolution NIR spectrograph CRIRES \citep{kaeufl04}, which is fed by the UT1 telescope of the Very Large Telescope facility. The cell is mounted inside the CRIRES ``warm optics'' box in an aluminum housing on a carriage that moves the cell in and out of the telescope beam. The cell sits just in front of the Nasmyth focus de-rotator in the converging f/15 beam from the telescope. At this location the cell is in front of all the spectrograph optics, as well as the instrument's integrated AO system. Observations of a star for radial velocity measurements are obtained with the cell in the beam, which causes the absorption lines of the cell to be imprinted on the stellar spectrum. The cell lines, whose position and shapes are well known, serve as a fiducial to precisely establish the wavelength scale and point spread function of the instrument at the time of the observation independently for each of the obtained spectra during analysis of the data (see \S4).

\begin{deluxetable*}{lcccccc}
\tabletypesize{\scriptsize}
\tablecolumns{7}
\tablewidth{0pc}
\tablecaption{Summary Of Observations}
\tablehead{
 \colhead{Star} &
 \colhead{exposure time (s)} &
 \colhead{$<$S/N pixel$^{-1}>$} &
 \colhead{exposures visit$^{-1}$} &
 \colhead{\# visits} &
 \colhead{\# epochs} &
 \colhead{date range}
}
\startdata
GJ\,551  & 20 & 160 & 12 & 27 & 4 & 2009.13 -- 2009.61 \\
GJ\,699  & 20 & 150 & 12 & 12 & 4 & 2009.14 -- 2009.61 \\
GJ\,406  & 60 & 140 & 8 & 11 & 3 & 2009.13 -- 2009.46 \\
GJ\,442B & 300 & 120 & 4 & 10 & 3 & 2009.13 -- 2009.46 \\
\enddata
\label{tab:data}
\end{deluxetable*}

The ammonia cell is not temperature stabilized during use because this is not possible given the existing space limitations in CRIRES. Therefore, the cell experiences temperature variations driven by the ambient temperature in the telescope dome during the observations. Its integration in the warm optics box provides a degree of thermal inertia, but does not prevent temperature variations on timescales of hours or longer. A temperature sensor is used to record the temperature inside the box during every science exposure. The median temperature during our observations over six months has been 8.8$\degr$\,C, with a standard deviation of 1.7$\degr$\,C, minimum of 5.9$\degr$\,C and maximum of 11.9$\degr$\,C. Simulations using calculated synthetic spectra for ammonia suggest that a temperature difference of 10$\degr$\,C would result in a systematic radial velocity shift of $\sim$\,1\,m\,s$^{-1}$. Our demonstration of 5\,m\,s$^{-1}$ long-term precision with data taken using the cell (see \S5) verifies that the temperature variations experienced by the cell are at least smaller than this level. Therefore, the ammonia cell enables high-precision calibration without the need for additional heating or temperature stabilization equipment. Indeed, the fact that ammonia is a gas at the temperatures experienced at ground-based observatories, and that its spectrum is relatively insensitive to temperature variations seems to give it an advantage over other possible gasses for NIR calibration cells.

We have made laboratory measurements of the cell using the Lund Observatory Bruker IFS125 HR FTS to obtain the characteristic spectrum that is needed for the radial velocity measurements during analysis of the data. This FTS has a full optical path difference of 4.8\,m, which gives a maximum resolving power of 10$^{6}$ at 5\,$\mu$m. We used a resolution of $\Delta k$ = 0.007\,cm$^{-1}$ ($R \equiv \lambda / \Delta \lambda \approx$ 620,000 at 2.3\,$\mu$m), fully resolving the cell's spectral absorption lines. The cell was illuminated with light from a 1,200.0\,$\degr$C black body source. Broad band filters were used to select the wavelength region containing the ammonia lines. The wavelength scale was set by the HeNe laser used in the sampling of the interferogram, and the absolute accuracy for this region is on the order of 0.001\,cm$^{-1}$. The measurements were obtained with the cell cooled to 13$\degr$\,C. This temperature is close enough to the temperatures the cell experiences while in use in CRIRES given the stability of ammonia as described above (i.e. $\Delta$T = 10$\degr$\,C corresponds to 1\,m\,s$^{-1}$). The resulting spectrum has a measured S/N $>$ 700 in the continuum near 2.3\,$\mu$m. In addition to this spectrum, we also obtained measurements of the cell at 24$\degr$\,C on two separate occasions 13 months apart. Comparison of these data reveal no change in the cell's spectrum, which supports our assumption of the cell's stability.

\section{OBSERVATIONS}
We have been using CRIRES with the ammonia cell for an ongoing planet search targeting very low-mass stars since February 2009. All observations so far have been performed in four Visitor Mode runs with lengths between four and six nights. We obtain observations of a portion of a single echelle order spanning the wavelength range 2292 -- 2350\,nm (a subset of the $K$-band), with three gaps of 3.0 -- 3.5\,nm that correspond to the spatial gaps between the four individual detectors in the mosaic used to record the data. The cell does not give enough suitable lines for calibration on the red-most detector. Therefore, only data from the three blue-most chips, which each contain enough ammonia lines, are utilized for the radial velocity measurements (total of 36.4\,nm spectral coverage). 

For most observations, including all those utilized for this paper, we use a spectrograph entrance slit with a projected width on the sky of 0\farcs2, which gives a nominal resolving power $R \approx$ 100,000. For the faintest stars we are observing ($K$ $>$ 9), a 0\farcs4 slit is used to obtain more photon counts at the cost of reduced spectral resolution. The CRIRES AO system is always utilized to improve the slit throughput. Multiple exposures (4 -- 16) are taken over a time-span of up to 20 minutes during each of the observational visits to a star. The telescope is nodded along the slit during a visit so that half the exposures are obtained in one nod position, and the other half in another position. We limit the exposure times to $\leq$300\,s in order to minimize the possible variations in the sky background between data taken in different nod positions. The exposure times are set when possible so that the average S/N will be $\sim$\,150\,pixel$^{-1}$. For the faint stars, the exposure time is simply set to its maximum value. An exposure time of 300\,s typically yields S/N = 100\,pixel$^{-1}$ for a late M dwarf having $K$=9.5. 

The obtained data are reduced and one dimensional spectra extracted using custom software that utilizes standard methods. Each chip is treated separately. First, the data are corrected for the signal arising from the bias voltage, dark current, and the amplifier glow by subtracting a dark frame taken with the same exposure time as the exposure in question. Multiple dark frames with the same exposure time are usually obtained during the course of an observing run. These are combined to create a master dark for the run using a sigma-clipped average, and this master dark is used for the correction. 

The next step in the reduction is the correction for the non-linear response of each individual pixel. The appropriate corrections as a function of the measured counts were determined from analyses of sequences of flat field images taken with different exposure times. The corrections determined from data obtained at different epochs over the last year are consistent, which suggests that the detectors do not exhibit a significant time-varying response.

After correcting for the non-linear response of the pixels, the data are then corrected for the non-uniform response of the pixels. The obtained flat field images used for this are first corrected for the bias, dark, and amplifier glow signals, and non-linearity in the same manner as for the science exposures. The multiple flat field exposures taken during a run are then combined to give a master flat field frame using a sigma-clipped average. All the science data for the corresponding run are divided by this master flat. Bad pixels are flagged during the creation of the master flat by identifying pixels whose sum over multiple exposures does not scale linearly with the exposure times.

Once the data have been corrected for the detector effects, the next steps are to subtract the background from the two dimensional images and extract one dimensional spectra. We do not utilize exposures obtained at different nod positions to estimate the background, although we have obtained data suitable for this procedure. Instead, the sky background for a given spectrum is estimated at each point in the dispersion direction using an average of the values of a number (20 to each side) of illuminated pixels that are more than 1\farcs5 from the peak of the image of the star in the spatial direction (note that CRIRES is a long-slit spectrograph and is not cross-dispersed). At theses separations the count levels are always much less than 1\% of the peak in the image of the star due to the use of the AO system to feed the spectrograph. We assume that the background level is the same for all pixels at the same point in the dispersion domain, which is a reasonable assumption despite the known curvature of the slit because there are no sharp sky emission lines in the wavelength range we observe in. The background as a function of spectral position estimated from this process is then subtracted from the data to complete the reduction of the two dimensional images.

We carried out tests of our method for estimating and subtracting the background compared to the method of simply subtracting an image taken in the opposing nod position. We found that a pixel-by-pixel subtraction of another science exposure leads to an increase in the noise consistent with that expected from the read noise of the detector. On the other hand, our method utilizes a number of pixels (typically around 40) to assess the background at each point in the dispersion domain, and thus introduces no significant noise while still delivering an accurate estimate. Therefore, this method yields slightly higher S/N in the end.

One dimensional spectra and their corresponding uncertainties are extracted from the two dimensional reduced data using an adaptation of the general optimal extraction algorithm described by \citet{horne86}. The background variance utilized for this algorithm is estimated as a function of spectral position during the calculation of the background level described above. This is usually slightly higher than expected from the known read noise of the detectors, but consistent with that estimated from propagating the uncertainties introduced at each step in the reduction. Pixels affected by cosmic ray strikes are identified and masked along with the previously identified bad pixels during the spectral extraction. The final S/N for the spectra utilized in this paper is typically 150\,pixel$^{-1}$. A summary of the observations for the data presented in this paper is given in Table~\ref{tab:data}. Included in the table are the nominal exposure times, typical S/N per pixel in the individual exposures, number of exposures obtained per visit, number of visits that have been performed, the number of epochs, and the time span of the observations.

\begin{figure}
\resizebox{\hsize}{!}{\includegraphics{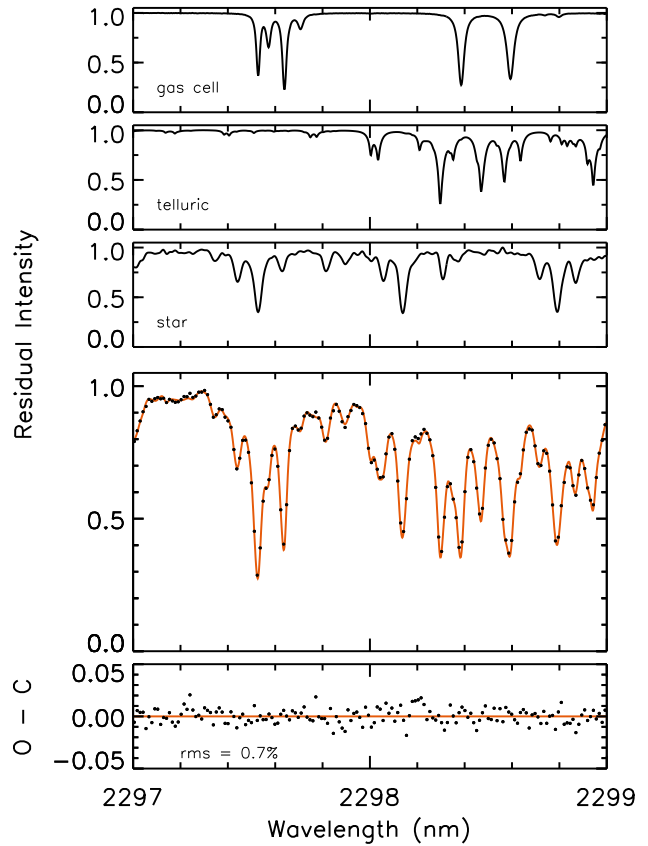}}
\caption{Example model components and fit for the radial velocity measurements. The components of the model are given in the top three panels: the spectrum of the ammonia cell (\textit{top}), synthetic telluric absorption spectrum (\textit{middle}), and the template stellar spectrum (\textit{bottom}). A comparison of a fit to the data is shown in the bottom two panels. (\textit{Top}) The observed spectrum (points) and the best-fit model (line). (\textit{Bottom}) The residuals from the fit (points). The short sections of spectra presented here represent $\sim$\,20\% of the full wavelength coverage of one CRIRES detector.}
\label{components}
\end{figure}

\section{RADIAL VELOCITY MEASUREMENT METHOD}

\subsection{Overview}
We measure the relative radial velocity of a star from a spectrum obtained with the ammonia cell using a method similar to the ``iodine cell'' method pioneered by \citet{butler96}. The basics of this method have been described extensively before \citep[e.g.][]{valenti95,butler96,endl00}. Therefore, we only give a brief description of the overall method in this subsection, and the details of the unique aspects of our implementation in the following subsections. We refer the reader to these previous papers for more information on the theory and techniques of the general gas cell method. 

The core of the gas cell radial velocity measurement algorithm is the fitting of a model spectrum to an observed spectrum to simultaneously calibrate it and determine the Doppler shift of the stellar lines. The model is a composite of the different components that appear in the observed spectrum. In our case, spectra of the gas cell, star, and telluric absorption.

The model for fitting the observed spectrum to determine the stellar radial velocity is constructed by combining the different components in to a single spectrum, and then convolving this spectrum with an instrumental profile (IP) and binning the result to the sampling of the observed spectrum. The free parameters in the model are adjusted to give the best match to the observed spectrum. The component spectra are combined in the model by sampling them on to the same wavelength scale and then multiplying them. Figure \ref{components} shows an example of the model components and a fit to an observed spectrum. Our model calculations (multiplication and convolution) are done on the observed pixel scale oversampled by a factor of six. The critical aspect for the radial velocity measurement is that the stellar spectrum is Doppler-shifted by an amount proportional to the radial velocity before the multiplication. This shift is one of the free parameters in the model, and its optimization to give the best-fit to the observed spectrum constitutes the measurement. 

\subsection{Template spectra}
Characteristic, or template, spectra of the three components that make up the model fitted to an observed spectrum during the radial velocity measurement process are required. The utilized spectrum of the gas cell is the FTS spectrum obtained at 13$\degr$\,C described in \S2. The telluric spectrum used in the model is based on spectrum synthesis following exactly the methods described by \citet{seifahrt09}. In summary, the template telluric spectra are synthesized using a radiative transfer code with a time-resolved (3\,hr resolution) model of the atmosphere over the observatory and an input line list. The radiative transfer code we use is LBLRTM\footnote{A description of the LBLRTM code and compilable source code are available at \url{http://rtweb.aer.com/lblrtm\_description.html}}. The input line list we use is based on the HITRAN database \citep{rothman09}. Initial tests using the HITRAN database as the input line list for the telluric spectrum synthesis resulted in poorer than expected fits to the observed spectra. This suggested that some of the HITRAN data could be modified so that the synthesized telluric spectra would more closely match the observed spectra, which would ultimately result in more precise radial velocity measurements. Therefore, we carried out work to determine improved data as described below.

We determined improved line data for the telluric spectrum synthesis by fitting synthesized spectra to a CRIRES spectrum of a telluric standard star (a fast rotating early B-type star) that was obtained with the ammonia cell. The empirical spectrum exhibited no lines attributable to the observed star, and contained only telluric absorption lines and ammonia lines from the cell. The ammonia lines in this spectrum served as a fiducial for the determination of the new telluric line data in a similar fashion as they are used for the radial velocity measurements. We fitted the segments of the empirical spectrum from the three utilized CRIRES chips separately. That is, we determined three unique line lists and imposed no constraints on the fitting linking the three lists to require, for example, the same methane or water abundance in the model atmosphere (see below), the same shift for lines that are part of the same band systems, or etc. In the fitting for each chip, we determined an overall wavenumber shift for all the telluric lines in the spectrum. We also determined individual wavenumbers and strengths for a total of 124 strong methane and water lines distributed fairly evenly over the three chips (i.e. the data for roughly 40 individual lines per chip were determined).

The fitting process to determine the new telluric line data was similar to the process for the radial velocity measurement process, but without the stellar spectrum because the telluric standard star star exhibited no lines in the wavelength regions we observe in. For the fit, the abundances of telluric water and methane used in the model were allowed to vary to improve the match to the haze of weaker lines that could not be reliably analyzed individually. The wavelength scale and IP (see description in \S4.3.1) were constrained by fitting the gas cell lines only. These lines were isolated by dividing the standard star spectrum taken with the gas cell by a spectrum of the same star taken immediately before without the gas cell. The typical adjustment to the wavenumbers of the telluric lines is on the order of 0.001\,cm$^{-1}$, which is consistent with the estimated uncertainties in the HITRAN database and the uncertainty in the absolute scale of the gas cell FTS spectrum. The line data determination was done once, and the resulting line list is used for all the radial velocity measurements.

The stellar template spectra we use for the radial velocity measurements are based on CRIRES spectra of the same stars. We determine a template spectrum uniquely for each star. The template for a given star is created from an observed spectrum using spectral deconvolution in a manner similar to the standard practice in the iodine cell method. However, a key difference for our ammonia cell NIR work arises due the presence of the telluric lines. 

The process to create a stellar template begins with the obtaining of a spectrum of the star in question without the gas cell. Immediately before or after this observation we also obtain a spectrum of a telluric standard star at a similar airmass (typically within 0.02) and also without the cell. The standard star spectrum, which contains only telluric lines, is modeled to determine the IP and wavelength scale at the time of the observation. The stellar template spectrum is also divided by the the standard star spectrum to remove the telluric lines. The IP determined from the analysis of the standard star spectrum is deconvolved from the stellar template spectrum after dividing out the telluric spectrum, and the determined wavelength scale from the analysis of the standard star is adopted.

\begin{figure}
\resizebox{\hsize}{!}{\includegraphics{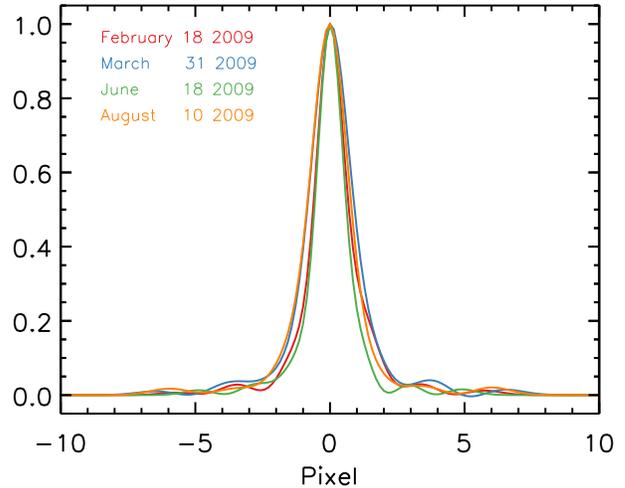}}
\caption{Example model IPs from each of the four observing runs when using a slit width of 0\farcs2.}
\label{ip}
\end{figure}

We do not utilize the gas cell in the process to create the stellar templates, which is in contrast to the typical technique for estimating the IP and wavelength scale of a template observation in the iodine cell method. The ammonia cell causes a different illumination of the spectrograph than when it is not in use because of a small tilt in the beam induced by the wedged windows. Therefore, the spectrograph IP and wavelength scale is different for observations with and without the cell. The stellar template spectra are obtained without the cell because the goal of these observations is to get as pure of spectra of the stars as possible, and so in theory the best calibration exposures for these data would also be obtained without the cell. The spectra of the telluric standard stars taken without the cell contain only telluric absorption lines, and these lines offer the possibility of calibration because we have already developed an appropriate technique for modeling them.

Tests indicate that using the telluric spectrum modeling described above on its own yields reasonable IPs, and a wavelength scale probably more closely matching that of the stellar template spectrum taken without the cell. Ultimately, the final radial velocities are insensitive to the choice of the exact method used for this step because the measured velocities are all relative and the absolute accuracy of the wavelength scale for the stellar template spectrum is not a limiting factor. 

\subsection{Model parameters}
In addition to the stellar radial velocity, the other free parameters in the model are the wavelength scale and IP of the observed spectrum, continuum normalization, water and methane abundances in the model atmosphere used for the telluric spectrum synthesis, and Doppler shift of the telluric spectrum. 

\subsubsection{Wavelength scale and instrumental profile}

The wavelength scale is parameterized by a second order polynomial. The IP model is constructed from the summation of 11 Gaussian profiles similar in style to a model originally suggested by \citet{valenti95} and described more recently by \citet{bean07}. In the IP model there is one central Gaussian, and five ``satellite'' Gaussians on both sides. The width of the central Gaussian and the heights of the satellites are the free parameters. The width and spacing of the satellites are fixed and set so that their half-widths just overlap. 

Some example IPs determined from the CRIRES data obtained with a 0\farcs2 slit width are shown in Figure \ref{ip}. The determined profiles are essentially Gaussians, but with some small deviations. The overall size and shape of the IPs also appears to vary with time. We believe both the deviations and time variability of the profiles are real. With regards to the first, the general morphology seen in the examples for different observing runs are consistent with the results from other spectra taken during the same run. The time variability is also observed when only a simple IP model consisting of a single Gaussian is used. Furthermore, the final measured radial velocities for known standard stars observed during our program (see \S5) are significantly degraded when we use only the simple IP model for the analysis. Therefore, the deviations and time variability do not seem to be an artifact of using the complex, multi-parameter model.

There are two reasonable explanations for the slightly non-Gaussian shape and time variability of the CRIRES IPs. First, using the AO system to boost the S/N often leads to under filling the entrance slit even when using a slit width of 0\farcs2. This leads to asymmetries in the IP. The AO performance varies with the atmospheric conditions and, therefore, the IP varies with the conditions as well. A second issue is that there were significant variations in CRIRES between the observing runs. For example, between the second and third run (May 2009) the instrument was warmed up and the detector mosaic was moved in the spatial direction so that it was centered on the slit. Also, a number of movable spectrograph components have poor position reproducibility (e.g. the entrance and intermediate slits, grating, and prism). The set point for the temperature for various parts of the instrument has been altered once as well. All these factors combine to suggest that indeed the CRIRES IP is slightly non-Gaussian and varying with time. The number of parameters (i.e. the number of satellites used) in our IP model was set so that the model captures these effects without over-fitting the data.

\begin{figure*}
\resizebox{\hsize}{!}{\includegraphics{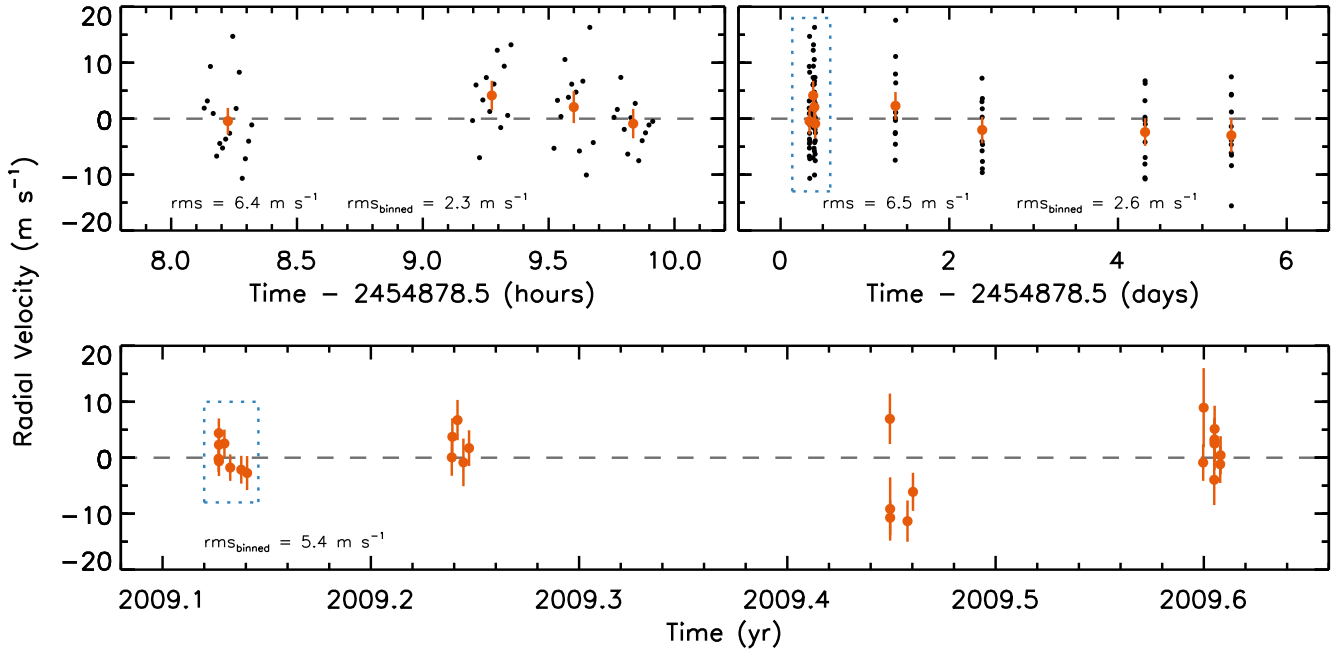}}
\caption{Measured radial velocities for Proxima Centauri. The black points are the results for individual exposures, and the orange points are the data binned over a given visit (always 12 individual exposures except for the first visit of 16 exposures). The error bars for the individual exposures are not shown for clarity. \textit{Top left} The results from a sequence of observations over two hours during one night. Another star was observed during the gap in the sequence. \textit{Top right} The results from the first observing run. The blue dotted box indicates the data shown in the top left panel. \textit{Bottom} The measured radial velocities obtained over six months (binned points only). The blue dotted box indicates the data shown in top right panel.}
\label{proxcen}
\end{figure*}

\subsubsection{Telluric Doppler shift}
In addition to the standard model parameters in the gas cell method for determining high-precision radial velocities, we found that a Doppler shift must also be applied to the telluric spectrum in our analyses. The possible need for such a shift was anticipated because the stability of the telluric lines at the few m\,s$^{-1}$ level was not known beforehand. However, the sizes of the shifts that came out of our analyses were unexpected. Over six months of observations we find that the analyses that yield constant radial velocities for known standard stars (see \S5) also yield telluric spectrum velocities that vary by up to a few hundred m\,s$^{-1}$ epoch to epoch (typical rms $\sim$ 75\,m\,s$^{-1}$).

We carried out two tests to investigate the origin of the apparent telluric line velocity variations. The first test was an analysis of spectra for telluric standard stars that were obtained with the gas cell regularly during our program so far. These spectra do not have stellar lines and may be modeled using only the gas cell template spectrum and a synthetic telluric spectrum. We analyzed the standard star spectra by carrying out the same kind of fitting as when determining the stellar radial velocity, but without that parameter. That is, we fitted the standard star spectra to solve for all the other model parameters described above except the stellar radial velocity. We analyzed 102 such spectra that were obtained over six months. We found that the obtained telluric velocity parameter was not varying at the same level as seen in the stellar radial velocity analysis, which is affected by the line blending. Instead, the determined telluric velocities have an rms of 15.8\,m\,s$^{-1}$. When the velocities were binned over each night, the rms is 7.3\,m\,s$^{-1}$. These are interesting findings in their own right, and we plan to publish a more detailed study of the variability of the telluric lines in the future. For now it is sufficient to know that this preliminary analysis shows that for Paranal Observatory, the telluric lines in our observing window are at least stable at a level of $\sim$\,20\,m\,s$^{-1}$ when they aren't confused with stellar lines.

The second test we carried out to check whether the telluric line velocity variability was real or not was to determine the stellar radial velocities using a synthetic stellar spectrum (generated with the PHOENIX code) for the template instead of the one determined empirically as described in \S4.2. The point of this test was that the synthetic stellar spectra do not have any possible residuals from telluric line division or deconvolution as the empirical templates do. This test yielded similar results for the telluric velocity parameter as the first test. 

The results from the two tests described above suggest that the telluric lines are much more stable than indicated by our main analysis to determine the stellar radial velocities. It is clear that the large variability of the telluric lines is only seen when using the empirical stellar templates. Therefore, we believe that the large telluric velocity shifts seen epoch to epoch come from imperfect telluric line division when creating the empirical templates. The imperfect telluric line division leaves residuals in the empirical stellar templates that are somewhat correlated with the telluric spectrum. These kinds of residuals are present despite our obtaining spectra of a standard star at a very similar airmass to the template observations for the telluric line removal. The residuals arise from the finite S/N of the observed spectra, and short timescale variations in the telluric absorption and the instrument.

The telluric residuals in the stellar template produce an effect on the telluric velocity shift parameter via cross-talk with the real telluric lines in observed spectra that are modeled for the radial velocity measurements. Over an observing season, the stellar lines move relative to the telluric lines due to the changing barycentric velocity of the observatory along the line of sight to the star (tens of km\,s$^{-1}$ peak-to-peak). Therefore, the stellar template in the model fit to the observed spectrum has to shift as well (this is what constitutes the measurement of the stellar radial velocity in the first pace). Therefore, the cross-talk between the template and observed spectrum requires that the telluric spectrum component in the model be shifted to absorb the correlated residuals during the fitting process. It is a subtle effect because the main drivers in the fitting are the match between the three component spectra in the model and their real corresponding lines in the observed spectrum. However, it is still necessary as either the telluric or stellar templates must be shifted to fit the data, and we have not been able to identify a way to filter out the effect in the fitting process.

Tests of the obtained precision indicate that using the empirical stellar templates gives significantly better results than synthetic templates despite the reduction in the degrees of freedom and the increased parameter uncertainties from correlations introduced by an additional free parameter (the telluric line shift). Despite being completely free of telluric residuals, the current state-of-the-art synthetic spectra for late M dwarfs do not sufficiently match the spectra of individual stars for high-precision radial velocity measurements. We are currently exploring the possibility of generating synthetic spectra specifically for this task, but it is uncertain whether such models will ever be more useful than the empirical templates despite the latter's flaws. Currently, synthetic stellar spectra might be useful as a sanity check if we see radial velocity variations in the measurements made using the empirical templates, but synthetic stellar spectra are not being used for the main velocity measurements. We have decided to use the empirical stellar templates and accept the telluric spectrum velocity parameter as a nuisance parameter for now. The issue of the telluric spectrum velocity parameter and its effect on the stellar radial velocities is taken up again in \S5.

\subsection{Fitting the data}
The parameters that yield the best fit between the model and the observed spectrum being analyzed are estimated using a standard non-linear least squares algorithm. The adopted uncertainty in the determined radial velocity is estimated from the covariance matrix and multiplied by the square root of the reduced chi-squared of the fit to account for the imperfect modeling. In addition to this error estimate, we also calculate the intrinsic radial velocity content of the data by considering the characteristics of the stellar spectrum (i.e. how sharp, deep, and numerous are the spectral lines), the obtainable level of calibration from the gas cell lines, and the S/N of the data. The intrinsic radial velocity content is the maximum level of precision that would be obtainable if the modeling process was perfect, and there was no covariance with other parameters - it is the photon-limited precision. We estimate this value using Eq.~6 from \citet{butler96} and calculate the obtainable precision for the stellar spectrum (i.e. the measurement precision) and the gas cell spectrum (i.e. the calibration precision). The results for the stellar spectrum and gas cell are added in quadrature to give an estimate of the intrinsic radial velocity content of the data. Our estimated uncertainties in the radial velocities are typically a factor of two larger than the calculated intrinsic content of the data.

The ammonia cell yields a spectrum similar in radial velocity quality to the spectrum of a slowly-rotating late M dwarf. Therefore, the calibration precision is on the same order as the Doppler shift measurement precision. Obviously this is not ideal. For example, iodine cells yield calibration precision a few orders of magnitude better than the intrinsic Doppler shift content of solar-type star spectra due to the very large numbers of lines molecular iodine exhibits in the utilized wavelength region. On the other hand, our situation is similar to that of using ThAr lamps for calibration in the visible where the calibration precision can be a limiting factor \citep{lovis07}. 

The spectra from each detector during a single exposure are analyzed separately (they are not further subdivided), and the final radial velocity for a given exposure is the weighted average of the results from the three detectors with usable data. The weightings are the estimated uncertainties from the fitting described above. The measured velocities are corrected for the barycentric motion of the observatory along the line of sight at the midpoint of the exposures to give the final values that may be used to monitor the motion of the target. All of the stars we are monitoring are very close by, and many have a large proper motion. Therefore, some exhibit a significant amount of secular acceleration. We calculate the expected trend for each star according the prescription of \citet{kurster03}, and subtract it when necessary.

\begin{figure}
\resizebox{\hsize}{!}{\includegraphics{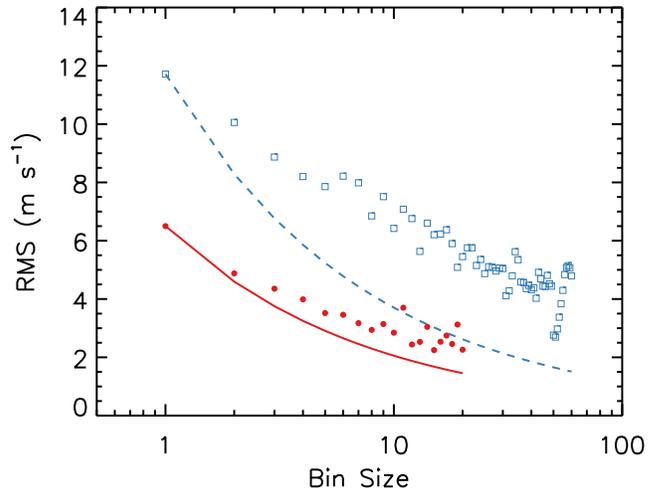}}
\caption{Residuals of the radial velocities for Proxima Centauri as a function of number of individual points binned. The circles are for the data taken during the first run (6 night span, 100 points total) and the solid line indicates the expected reduction in rms ($N^{-1/2}$ where $N$ is the number of data points binned) normalized to the unbinned (N=1) rms. The squares are for all of the obtained data, and the dashed line corresponds to the expected reduction in rms for those data.}
\label{binned}
\end{figure}

\section{VELOCITY PRECISION TESTS}
In addition to the very low-mass stars we are monitoring as part of our ongoing planet search, we have also frequently observed Proxima Centauri (GJ\,551) and Barnard's Star (GJ\,699) to test our observational and data analysis methodology. These stars are two of the few very low-mass stars for which it is possible to obtain high-precision radial velocity measurements in the visible wavelength range (due to their being very nearby, and thus bright even at visible wavelengths), and previous work has shown them to be radial velocity constant at the level of 3\,m\,s$^{-1}$ \citep{kurster03, endl08, zechmeister09}. We also present here data from two of our planet search targets that so far appear constant. 

\subsection{Results for Proxima Centauri}
Figure \ref{proxcen} shows the radial velocities measured for Proxima Centauri from spectra obtained during 17 nights over six months. These velocities are completely consistent with the previous findings that the star does not exhibit large radial velocity variations. The rms of the radial velocities measured for 100 individual spectra obtained over a six night observing campaign is 6.5\,m\,s$^{-1}$ (see the top right panel in Figure \ref{proxcen}). The intrinsic radial velocity content of the data (i.e. the photon-limited precision) is typically 5.6\,m\,s$^{-1}$, while our estimated errors are typically 9.1\,m\,s$^{-1}$. When the data are binned over complete visits (usually 12 individual exposures of 550\,s total observing time) during this observing run, the velocities exhibit a dispersion of 2.6\,m\,s$^{-1}$. The unbinned and binned velocities exhibit the same level of dispersion from timescales of two hours (see the top left panel in Figure \ref{proxcen}) up to six days. The expected dispersion in the binned data from the intrinsic radial velocity content is 1.6\,m\,s$^{-1}$, while that expected from our estimated errors is 2.6\,m\,s$^{-1}$. 

The rms of the individual data points over short timescales is 30\% smaller than our estimated errors, but is similar to the intrinsic quality of the data (i.e. that expected from photon statistics). This suggests that our error estimation method might be slightly overestimating the uncertainties on this timescale. The main difference between our estimated errors and the photon limited precision is that we also take into account the imperfect modeling by adjusting the nominal errors according to the chi-squared of the fit to the observed spectrum. The main limiting factor in the modeling is the stellar template, which is not optimal because it is created through a deconvolution of a finite S/N spectrum. Spectral deconvolution in general amplifies noise and can not yield the exact underlying spectrum. Also, as discussed above, the empirical templates contain residuals from the telluric line division. Other secondary factors possibly leading to imperfect reproduction of the observed spectra are the limitations of the IP model and the synthesized telluric spectra.

\begin{figure*}
\resizebox{\hsize}{!}{\includegraphics{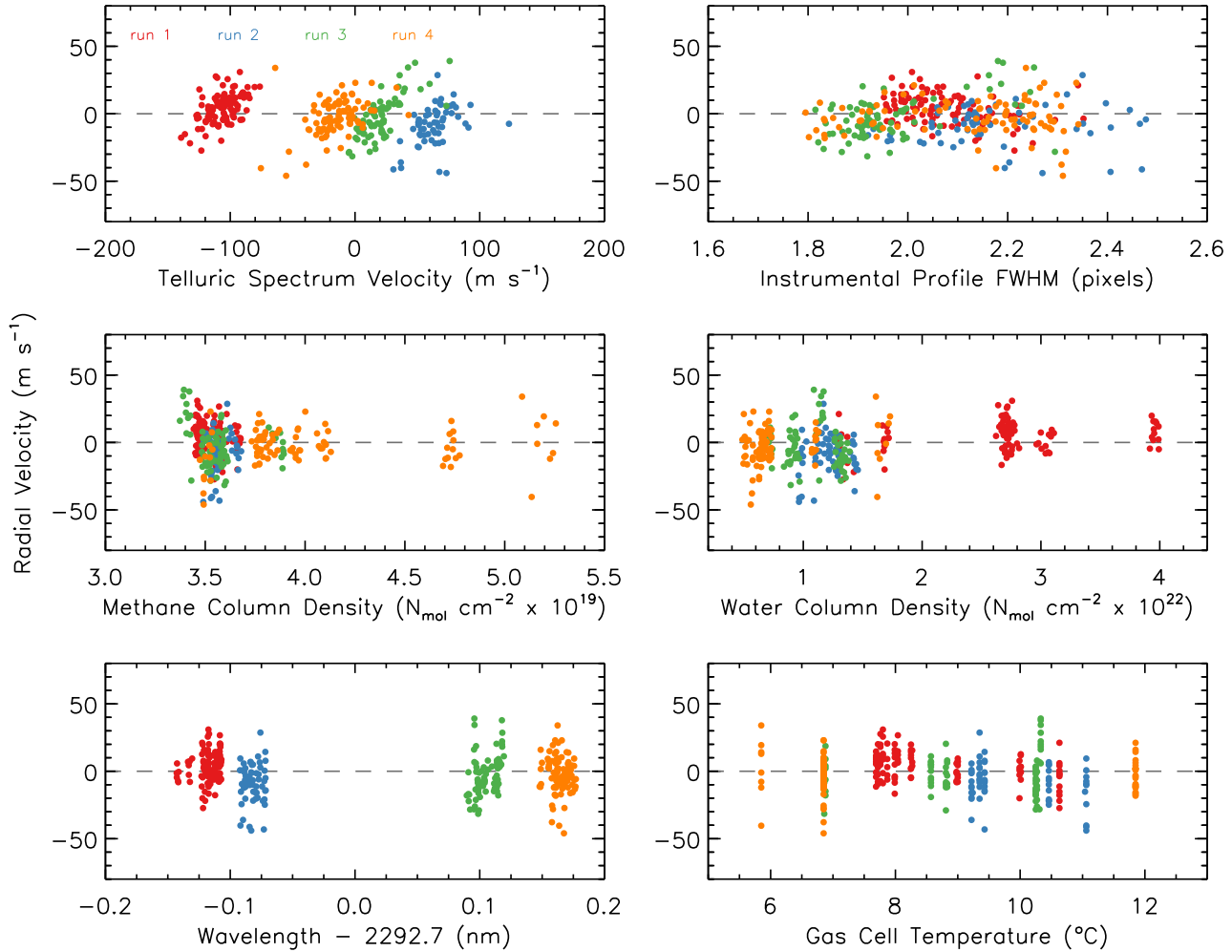}}
\caption{Radial velocities for Proxima Centauri measured from the spectra on the blue-most CRIRES detector against some of the free parameters in the model used to fit the observed spectra. The colors of the points correspond to the different observing runs as indicated in the top left panel.}
\label{param}
\end{figure*}

Over a short time-span the systematics arising from the imperfect modeling will be same because the IP, wavelength scale, telluric spectrum, and etc. will be similar for all the data. Therefore, all the measurements will be similarly affected and the systematics will be reduced. However, the rms of the binned data over a few nights is slightly higher than expected from the photon limited precision, but consistent with our estimated errors. As the data are binned, and the expected precision drops to a few m\,s$^{-1}$, the imperfect modeling probably becomes an issue again. 

The dispersion in the radial velocities measured for Proxima Centauri over the full monitoring time-span (see the bottom panel in Figure \ref{proxcen}) is 11.7\,m\,s$^{-1}$ for the individual spectra (not shown in Figure \ref{proxcen}) and 5.4\,m\,s$^{-1}$ for the binned data. The typical value for our estimated errors over the full time-span is 11.5\,m\,s$^{-1}$, and the expected dispersion in the binned data from these errors is 3.4\,m\,s$^{-1}$. The typical photon-limited error is 7.0\,m\,s$^{-1}$, and the expected dispersion in the binned data from these errors is 2.0\,m\,s$^{-1}$. 

The dispersion in the results for the individual spectra over long timescales is larger than that expected from the intrinsic radial velocity content of the data. However, the dispersion is consistent with our estimated errors and this suggests that the imperfect modeling is important over long timescales, which is consistent with the discussion above. However, the rms of the binned data over six months is 60\% higher than expected from the estimated errors, which indicates that an additional source of noise is present on long-term timescales. We attribute this additional noise to the cross-talk between the telluric lines in the spectra and residuals from the telluric line division in the stellar template used to model the spectra in the determination of the radial velocities as described in \S4.3.2.

Figure \ref{binned} shows how the rms of the Proxima Centauri radial velocities depends on the number of points binned. Over short timescales (i.e. a single observing run, circles in the figure) the rms of the data nearly follows the expected trend (i.e. the rms goes as $N^{-1/2}$, where $N$ is the number of data points binned - see the solid line in the figure). The noise floor over short timescales appears to be $\sim$\,2\,m\,s$^{-1}$. It is worth noting that although Proxima Centauri is observable in the visible, the previously published data only indicate that it is at least stable at the level of 3\,m\,s$^{-1}$. It could be that the star is variable at a lower level, and 2\,m\,s$^{-1}$ is potentially only an upper limit to the noise floor of our method over short timescales. 

Over long timescales (i.e. an entire observing season), the rms of the Proxima Centauri radial velocities is a factor of two larger than on short timescales, and the data show significant deviations from the expected reduction in rms with binning (squares and dashed line in Figure \ref{binned}). We attribute this to the additional noise arising from the telluric cross-talk described above. The noise floor over long timescales appears to be $\sim$\,4\,m\,s$^{-1}$ from these data, which is consistent with the results seen for other stars (see below).

In Figure \ref{param} we plot the radial velocities for Proxima Centauri measured from the spectra on the blue-most CRIRES detector against some of the free parameters in the model used to fit the observed spectra. This figure demonstrates the typical range for the parameters, and allows a check for correlations. We find that the radial velocities show no correlation with the IP width, methane or water abundance in the model atmosphere used for the telluric spectrum synthesis (shown as the column density of these species), wavelength zero point, or temperature of the gas cell. Within a given run (epoch), the stellar radial velocities are correlated nearly one-to-one with the telluric spectrum velocities. This is expected because the variability of the telluric spectrum velocity parameter on the order of a few hundred m\,s$^{-1}$ arises from changes in the stellar velocity shift on the order of a few km\,s$^{-1}$. Within a given run the barycentric velocity of the observatory along the line of sight will vary by at most a few tens of m\,s$^{-1}$ during all the observations of a given star. The telluric residual cross-talk causes the telluric lines to be dragged with the stellar template spectrum when it is shifted in the model. Within a given run, the shift correlation is just essentially one-to-one because of the small scales involved.

As can be seen in Figure \ref{param}, the stellar radial velocities are not correlated with the telluric velocities over the four epochs that the data presented here were obtained. The key point is that freeing the telluric velocity parameter yields constant radial velocities for the known constant stars over long-term timescales. This validates our use of this parameter. However, the telluric lines in general limit the obtainable precision, and the need for this shift parameter is an aspect of that.

\begin{figure}
\resizebox{\hsize}{!}{\includegraphics{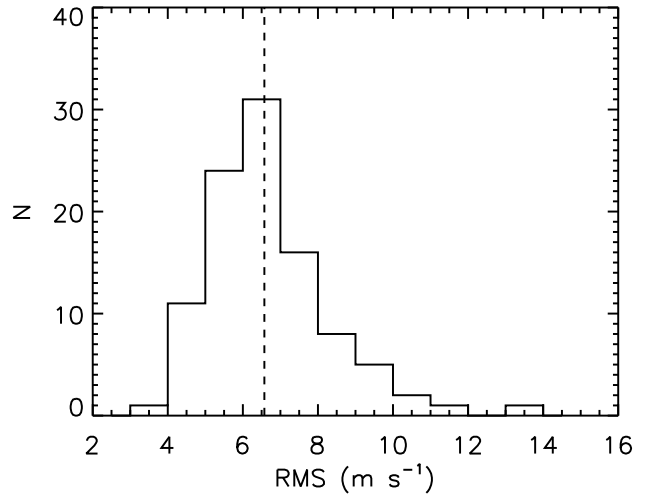}}
\caption{Histogram of the rms residuals for the simulation of precision obtainable with four radial velocity measurements per night for Proxima Centauri. The dashed line gives the median of the sample, which is 6.4\,m\,s$^{-1}$.}
\label{simul}
\end{figure}

Proxima Centauri (K=4.4) is very bright relative to the stars we are targeting for the planet search (K = 7 -- 10). Therefore, we were able to obtain many more spectra of it during each visit (typically 12 spectra in less than 10 minutes) than we are obtaining for most of the other stars. Also, we specifically obtained a large number of observations of Proxima Centauri to carry out the velocity precision tests presented here. To investigate what level of precision the Proxima Centauri data suggest for the stars in our program for which we are obtaining fewer spectra per visit, we used a simulation.

We typically obtain four spectra per visit for the planet search targets with a similar S/N as the Proxima Centauri spectra. For the simulation, we randomly drew and binned four radial velocity measurements from the Proxima Centauri data for each night of observation to create an example measurement sequence. The rms of each of the trial sequences was calculated to estimate the obtained precision. We repeated the simulation 100 times and a histogram of the resulting rms values are plotted in Figure \ref{simul}. The median rms of the 100 trials is 6.4\,m\,s$^{-1}$. This is likely a good estimate of the obtainable precision for a realistic set of observations on our planet search targets.

\begin{figure}
\resizebox{\hsize}{!}{\includegraphics{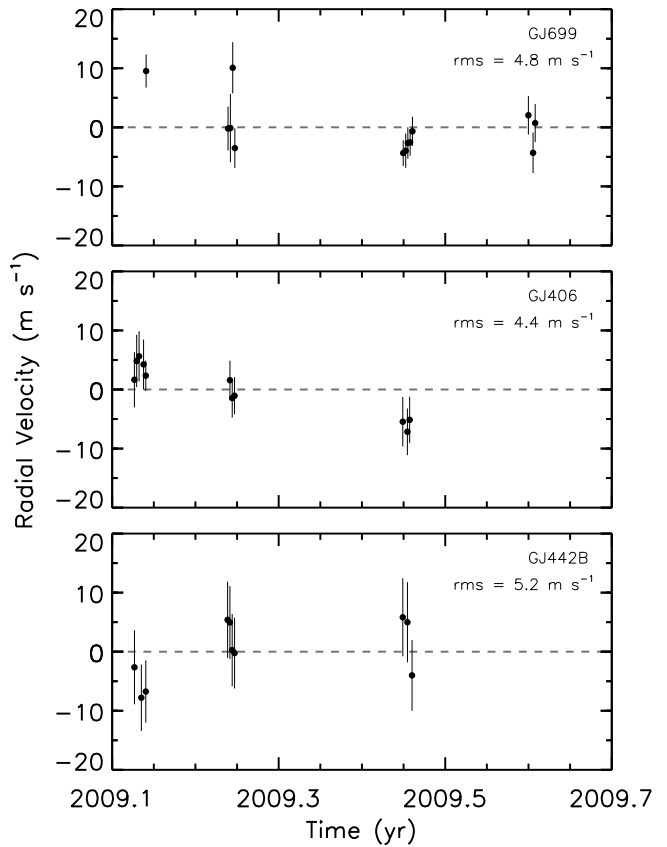}}
\caption{Radial velocities measured for Barnard's Star (\textit{top}), CN Leo (GJ\,406, \textit{middle}), and GJ\,442B (\textit{bottom}).}
\label{others}
\end{figure}

\subsection{Results for Barnard's Star and others}
In Figure \ref{others} we plot the radial velocities we have measured for the late M dwarfs Barnard's Star, CN Leo (GJ\,406), and GJ\,442B over time spans of four to six months. As mentioned above, Barnard's Star is a known radial velocity constant star \citep[minus its significant secular acceleration,][]{kurster03} at the level of 3\,m\,s$^{-1}$. \citet{reiners09a} presented radial velocities measured from visible spectra for CN Leo obtained over three consecutive nights. These data were strongly affected by a large flare that occurred during the observations, but outside this event the radial velocities exhibit a dispersion of $<$\,10\,m\,s$^{-1}$. No high-precision radial velocities have been published for either CN Leo or GJ\,442B over long timescales. CN Leo and GJ\,442B are typical targets in our planet search and we present data here only as an illustration of our obtainable precision. Future work will focus on detailed interpretation of the results in the context of the planet search.

The radial velocities for all three stars shown in Figure \ref{others} exhibit dispersions on the order of 5\,m\,s$^{-1}$. Eight to twelve measurements per visit were obtained and binned for Barnard's Star and CN Leo. We performed simulations using the data for these stars to investigate the precision obtainable with only four measurements per visit as we did for Proxima Centauri. In both cases we obtained typical dispersions of 4 -- 6 \,m\,s$^{-1}$.

The star GJ\,442B is very similar in characteristics to the main planet search targets we are following and the usual four measurements per visit were obtained and combined. This star has V=15.4 and, therefore, would be a very challenging target for obtaining high-precision radial velocities with existing visible wavelength spectrographs. However, as an M5 dwarf it is very red and it has K=8.0. With these data we have demonstrated that by using CRIRES at the VLT with our new gas cell we can obtain measurement precisions of $\sim$\,5\,m\,s$^{-1}$ in 20 minutes total exposure time on such a star. 

\section{DISCUSSION}
The general technique of measuring radial velocities from NIR spectra has recently been receiving a lot of attention as a possible important direction for the observational study of exoplanets for the reasons described in \S1.1 \citep[e.g.][]{lunine08}. However, up to now the feasibility of obtaining NIR radial velocity precision similar to that routinely obtained in the visible was largely unknown due to uncertainty about a number of technical issues. The main open question has been how to calibrate NIR spectra, with a second issue being whether the quality of NIR detectors is sufficient.

We have described a new gas cell suitable for calibrating spectra of cool dwarfs in the $K$-band, and demonstrated long-term radial velocity precisions of $\sim$\,5\,m\,s$^{-1}$ from data obtained with CRIRES at the VLT when using the cell. It is worth noting that the CRIRES detectors have severe cosmetic blemishes relative to CCDs. However, the effects seem to be stable and we have been able to calibrate them out to an acceptable level. Furthermore, the detectors were not even the highest-quality NIR detectors that were available at the time CRIRES was built \citep[ca. 2002,][]{dorn04}, and are certainly not competitive with the best detectors available as of this writing. Therefore, the issue of the quality of NIR detectors requires attention, but will not likely be a limiting factor for new instruments aiming at $\sim$\,1\,m\,s$^{-1}$ NIR radial velocity precision.

Our ammonia gas cell has proven to be useful for NIR calibration down to a few m\,s$^{-1}$ precision despite not being temperature stabilized. Therefore, it could be useful for high-precision radial velocity measurements with other instruments not originally designed for excellent stability. The precision afforded by the cell and the measurement method we have described is already an order of magnitude better than the best previously obtained long-term NIR precision on a star other than the Sun. As a result, it now enables the possibility of searching for planets around a much larger number of very low-mass stars than was feasible before. For example, our planet search project includes 31 objects with estimated masses below 0.2\,M$_{\odot}$, 22 of which have estimated masses below 0.15\,M$_{\odot}$. A previous high-precision radial velocity planet search utilizing a visible wavelength spectrograph on an identical telescope (UVES on the UT2 telescope of the VLT) was only able to target two objects with estimated masses below 0.2\,M$_{\odot}$, and only one of these two has an estimated mass below 0.15\,M$_{\odot}$ \citep{zechmeister09}.

In addition to enabling the search for planets around more low-mass stars, the gas cell and radial velocity measurement algorithm we have developed also opens up a new frontier on the search for potentially habitable planets. The orbital period range for a planet in the habitable zone around a star with a mass M = 0.10\,M$_{\odot}$ would be 3 -- 21 days \citep{selsis07}. The 5\,m\,s$^{-1}$ precision obtainable with our method corresponds to the velocity semi-amplitudes induced by a 2.5\,M$_{\oplus}$ planet or a 4.6\,M$_{\oplus}$ planet on the inner and outer edges of the habitable zone respectively. Therefore, it should be possible to detect Super-Earth type planets in the habitable zones of very low-mass stars with a reasonable expenditure of observing time on current facilities.

Altogether our results have shown that obtaining high-precision NIR radial velocities is possible, and we see no technical reason why a level of precision of 1\,m\,s$^{-1}$ could not be achievable with a highly stabilized NIR grating spectrograph. However, the design of such a next generation instrument will have to carefully weigh the availability of calibration against the availability of information (i.e. spectral lines) from the stars being targeted like we have done with our observing program.

As discussed above, we are carrying out a planet search using CRIRES with the ammonia cell and this paper is the first in a planned series from the project. The next papers in the series will present high-precision NIR radial velocities of our M dwarf planet search targets. The data will allow us to detect planets around these stars with similar masses and orbital separations as those that can be detected around solar-type stars using state-of-the-art visible wavelength radial velocities.

\acknowledgments
We thank the ESO staff, in particular Alain Smette and Hans-Ulli K\"{a}ufl, for their assistance with the implementation of the gas cell in CRIRES and carrying out our observations. This work was partly supported by the DFG through grants GRK 1351 and RE 1664/4-1, and the BMBF through program 05A0GU2. The Lund IR-FTS was financed through  a grant from the Knut and Alice Wallenberg Foundation. J.L.B. acknowledges research funding from the European Commission’s Seventh Framework Program as an International Incoming Fellow (grant no.~PIFF-GA-2009-234866). A.S. acknowledges financial support from the NSF under grant AST-0708074. H.H. acknowledges funding from the Swedish Research Council (VR). H.N. acknowledges the financial support from the Lund Laser Center through a Linnaeus grant from the Swedish Research Council (VR). A.R. received support from the DFG as an Emmy Noether Fellow.

{\it Facilities:} \facility{VLT:Antu (CRIRES)}

\bibliographystyle{apj}
\bibliography{ms.bib}

\end{document}